\documentstyle[12pt,aasmanuscript,sub]{article}
\newcommand{\be}{\begin{equation}}
\newcommand{\ee}{\end{equation}}
\begin{document}
\title{A Parallel Tree Code}
\author{John Dubinski}
\affil{Board of Studies in Astronomy and Astrophysics\break
         University of California, Santa Cruz, CA 95064\break
         dubinski@ucolick.org}
\markboth{DUBINSKI}{PARALLEL TREE CODE}

\begin{abstract}

We describe a new implementation of a parallel N-body tree code.  
The code is load-balanced using 
the method of orthogonal recursive bisection to subdivide the N-body system
into independent rectangular volumes each of which is mapped to a processor
on a parallel computer.
On the Cray T3D, the load balance in the range  of 70-90\% depending on the
problem size and number of processors.
The code can handle simulations with $>$ 10 million particles roughly a
factor of 10 greater than allowed on vectorized tree codes.
\end{abstract}

\section{Introduction}

During the past decade, $N$-body tree codes have been applied successfully 
to various problems in galaxy dynamics, galaxy formation and
cosmological structure formation 
(Barnes \& Hut 1986; Hernquist 1987; Dubinski 1988).
Computing time only scales as $N\log N$
so they provide a relatively fast way to solve the general collisionless N-body
problem.  
For the most part, the favored tree code in studies has been the Barnes-Hut 
(BH) code (1986) mainly because it was widely distributed by its
authors but also because it has been easy to implement.
Different tree-based algorithms have also been used successfully for
various problems (Appel 1985; Jernigan \& Porter 1989; 
1993; Benz, Bowers, Cameron \& Press 1990; Steinmetz \& Muller 1993).
Tree codes have also been linked up with smoothed particle hydrodynamics
(SPH) (Hernquist \& Katz 1989; Benz et al 1990; Steinmetz \& Muller 1993).
Tree code simulations running 
on the latest workstations and vector supercomputers
are generally restricted to $N < 10^6$ because of memory and time limitations.
Cosmological and hydrodynamical simulations require ever greater dynamic
range, so there is a strong desire to increase these limits. 

During the past few years, there has been a paradigm shift in
supercomputing with the movement from vector machines
to the new massively parallel machines.  Parallel supercomputers
are collections of hundreds to thousands 
of independent smaller commodity processors
interconnected with an internal high speed communication network.
The NSF supercomputing centers support several machines: the Connection Machine 5,
(CM-5), the Intel Paragon,
the IBM SP/2, and the Cray T3D.
Each processor
operates independently but communication software allows messages containing
data to be exchanged rapidly with other processors.  
In principle, a problem can be partitioned among the $N$ processors and a
maximum $N$-fold increase in computational speed can be realized.
In practice, the speed up is smaller because of the extra time needed for
exchanging data in message-passing algorithms.

Parallel machines offer a new route to very fast computation 
if algorithms can be
redesigned to conform to the message-passing paradigm and communication 
can be minimized.
The new algorithms are often very different than their sequential
counterparts and require a considerable effort to redesign.
The key to a successful algorithm is good {\em load balance}: both data
and computational work must be distributed evenly among the processors.
A common way of achieving load balance
on parallel machines is through {\em domain
decomposition} 
(Fox, Williams \& Messina 1994).
The physical domain of the problem 
is partitioned into smaller subdomains and the physical quantities
of these subdomains: either values of density, pressure and temperature at
grid elements in an Eulerian fluid code or collections of particles and
their attributes
in an N-body code are assigned to each processor.
The trick in designing a parallel algorithm is 
finding a way of partitioning the domain so that data and work are evenly
distributed and communication is minimized. 

There have been many efforts to parallelize the BH tree code.
Hillis and Barnes (1987) and
Makino and Hut (1989) ported the code to the Connection Machine-2 
but at the time there was little gain over the existing vector machines.
Salmon (1990) introduced a new parallel tree algorithm 
using a domain decomposition based on the orthogonal
recursive bisection of the N-body volume into
rectangular subvolumes for purposes of load balance.
The code has been applied to dark halo formation with
initial  density fluctuations based on different power spectra
(Warren et al. 1992; Warren 1994; Zurek et al. 1994).
Warren \& Salmon (1993) 
have also recently designed a new algorithm which uses a
load-balancing scheme based on a parallel hashed oct-tree 
(also Warren 1994).
Dikaiakos and Stadel (1995) also have developed another variant of Salmon's
algorithm in their parallel tree code PKDGRAV.
Parallel N-body tree codes have also been discussed as a pure 
computer science problem
(Bhatt et al. 1992; Pal Singh 1993; Pal Singh et al. 1995).

In this paper, we present a new version of Salmon's (1990) parallel
tree algorithm.
In \S 2, we describe the BH tree code as implemented at the node level
in the parallel code.
In \S 3, we describe a new portable
implementation of Salmon's parallel algorithm using the Message Passing
Interface (MPI) package to handle processor communication.
In \S 4, we examine the code's performance on the Cray T3D,
checking the accuracy, speed, and load balance.
The term ``node'' is often used to refer to a single processor on a parallel
computer but it is also used to refer to a data structure in a tree.
To avoid confusion, I will only use the term ``node'' 
in reference to tree structures and ``processor'' for a computational ``node''.

\section{Tree Codes}

The N-body problem involves advancing the trajectories of $N$ particles
according to their time evolving mutual gravitational field.
In the simplest algorithm, the force on each particle is determined by
direct summation of the contributions from all of the N-1 neighbours.
In a discrete integration,
the forces at each timestep are then used to
advance the particles along their trajectories according to a numerical
integration scheme such as the leapfrog method.
Computational costs of direct summation scale as $O(N^2)$ making this
algorithm expensive.  In collisionless systems like galaxies, however,
one can tolerate small errors in the force for improved performance by using
techniques which approximate the gravitational field.
Tree codes are one class of methods which accomplish this and have the
advantage of scaling only as $O(N\log N)$ in computational cost (Appel
1985; Barnes \& Hut 1986; Jernigan \& Porter 1989).

The essence of a tree code is the recognition that 
the gravitational potential of a distant group of particles can be
well-approximated by a low-order multipole expansion.
A set of particles can be arranged in a hierarchical system of groups in
the form of a tree structure.
The entire set can be subdivided into several groups and each of these
groups can be broken down in succession in the hierarchy until the groups
contain at most 1 particle.
There are many different methods for
grouping particles hierarchically in this way (Appel 1985; Jernigan \&
Porter 1989; Barnes \& Hut 1986).
The evaluation of the potential at a point reduces to a
descent through the tree.
One sets
a minimum distance a point can be from a group to use a multipole
expansion.
If the point is sufficiently distant from a group, 
the multipole expansion is used to find the
potential from that group, 
while if the point is too close to the group, each of its child subgroups
are examined.  The procedure continues until all groups are broken down as
far as they need be.

\subsection{The Barnes-Hut Tree Code}

The Barnes-Hut (1986) algorithm works by grouping particles
using a hierarchy of cubes arranged in oct-tree structure i.e. each node in
the tree has 8 siblings.
The system is first surrounded by a single cube or cell encompassing all of the
particles.
This main cell is subdivided into 8 subcells,
each containing their own subset of the particles.
The tree structure continues down in scale until 
cells contain only 1 particle.
For each cell or node in the tree, we calculate
the total mass, center of mass and higher order multipole moments (typically
only up to quadrupole order).
This tree structure can be built very rapidly making it feasible to rebuild
it at each time step.
The tree can be constructed bottom-up i.e. by inserting one particle at a
time (Barnes \& Hut 1986) or top-down by sorting particles across divisions.
Both methods take $O(N\log N)$ time and in
practice only a few percent of total time per step.

The force on a particle in the system can be evaluated by
``walking'' down the tree level by level beginning  with the top cell.
At each level, a cell is added to an interaction list
if the cell is distant enough for a force evaluation.
If the cell is too close, it is ``opened'' and the 8
subcells are either used for the force evaluation or opened further.
The walk ends when all cells which pass the opening test and any single
particles are acquired.
The accumulated list of interacting cells and particles is 
then looped through to
calculate the force on the given particle and this amounts 
to the bulk of the computation.
In this way, the number of interactions computed is significantly smaller
than a direct N-body method  with the number scaling as $O(N\log N)$ (Barnes
\& Hut 1986).
Typically, there are $\sim 1000$ interactions per particle on average in
a simulation with $10^6$ particles making the algorithm significantly factor
than a direct summation method.

\subsection{Cell Opening Criterion}
\begin{figure}[t]
\centerline{\epsfbox{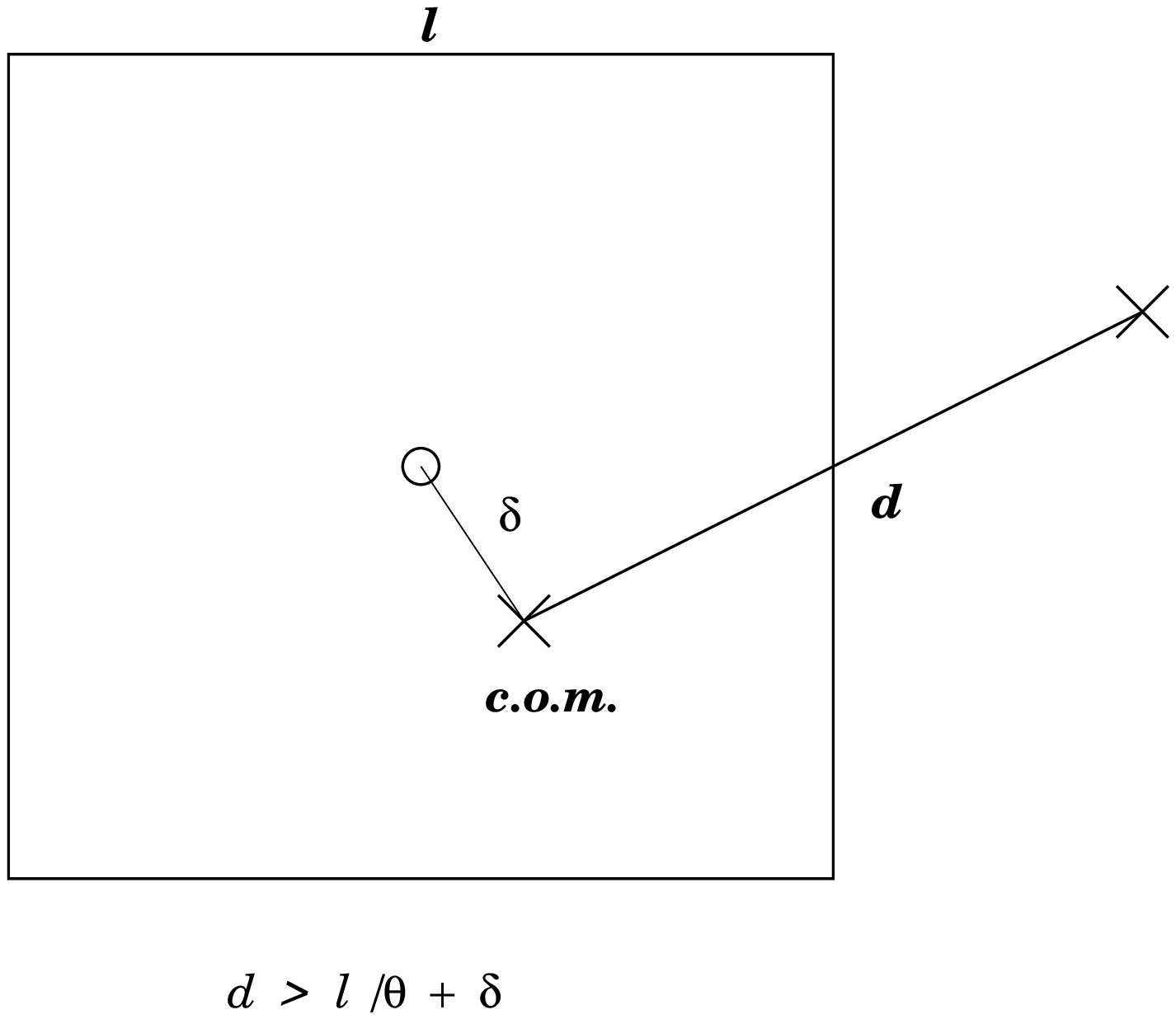}}
\caption{Geometry of the Barnes' new cell-opening criterion used 
in the parallel tree code.}
\label{fig-bhopen}
\end{figure}

There are various criteria 
for determining whether a cell is sufficiently distant for a force
evaluation.  
The simplest criterion 
was introduced by Barnes \& Hut (1986) based on an
opening angle parameter, $\theta$.  If the size of a cell is $l$ and the
distance of the particle from the cell center of mass is $d$, we accept the
cell for a force evaluation if 
\begin{equation}
d > l/\theta.
\end{equation}
Smaller values of $\theta$ lead to more cell openings and more accurate
forces.
Typically, $\theta=1$ gives accelerations with errors around 1\%
(Hernquist 1987).
Salmon \& Warren (1994) showed, however, that this criterion for calculating
the potential can cause gross errors in some pathological cases in which
the center of mass is near the edge of the cell.
They introduced various alternatives which avoid this problem as 
did Barnes (1994).
After some experimentation with various opening criteria,
we have adopted Barnes' (1994) modification which circumvents this problem.
The new criterion for cell opening is 
\begin{equation}
d > \frac{l}{\theta} + \delta,
\label{eq-bhopen}
\end{equation}
where $\delta$ is the distance between the center of mass
of a cell and its geometrical center (Figure \ref{fig-bhopen}).
This criterion guarantees that if the center of mass is near the cell edge
only positions removed by an extra distance $\delta$ use the cell for a
force evaluation, while if
the center of mass is near the cell center it reverts to the old criterion.

\subsection{Grouping}
Tree walks themselves include a sizeable overhead in computation since the
cell opening criterion must be computed many times for each particle.
Barnes (1990) 
introduced one final improvement to reduce the number of tree walks
for a net gain in computing speed.  
The strategy is to find all cells in the tree which contain less than
$N_{crit}$ particles where typically $N_{crit} \sim 32$.
The tree walk proceeds as before but instead of defining
$d$ as the particle distance from a target cell's center of mass, $d$ is 
defined
as the distance between the nearest edge of the cube encompassing the group
to the target cell's center of mass.
The accumulated interaction list is then used for all of the particles in a
given group and the forces are guaranteed to be at least as accurate as
those of an individual particle at the edge of the group's cube.
In practice, this can lead to a considerable speedup for a fixed force
accuracy (Barnes 1990).

\subsection{Non-recursive tree walks}

When tree structures are part of 
 an algorithm,
it is natural to program functions recursively.
Unfortunately, there can be
considerable overhead from recursive calls.
The
tree walk is called many times at each timestep and so it is best
to eliminate recursion in this function.
Non-recursive tree walks can be
accomplished by arranging the tree nodes in a linked list (Dubinski 1988;
Makino 1989).
Each node is set up to contain 
a pointer to its first child and adjacent sibling.
If a node is the last child in a level,
its sibling pointer is assigned to its parent's sibling.
A node which contains 1 particle will have only one adjacent sibling 
and no children.

\begin{figure}[t]
\centerline{\epsfbox{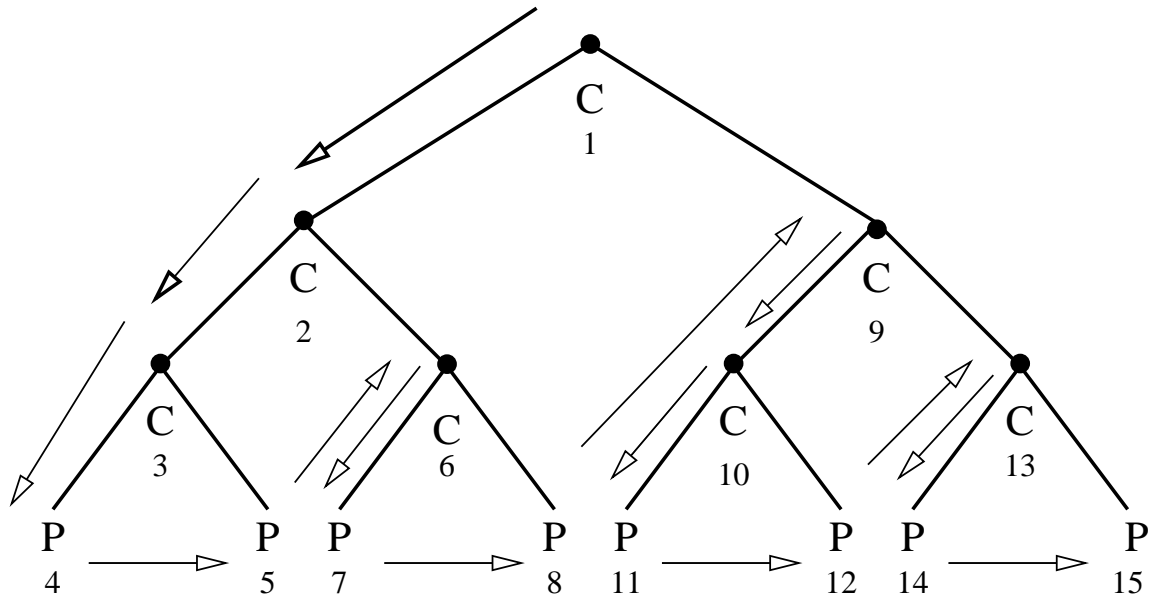}}
\caption{Idealized binary tree showing the order of nodes for a
non-recursive tree walk.  Cell nodes are labelled with ``C'' and particle
nodes with ``P''.  The index of the node in the list is also shown.}
\label{fig-nonrecur}
\end{figure}

The nodes are resorted as follows (Figure \ref{fig-nonrecur}).
The first node in list is set to the root cell containing all of the particles.
The descent of the tree always proceeds to the current node's first child
which is then added to the list.
If the node contains only one particle the walk proceeds to the node's
sibling which is then added to the list.
Once the nodes are sorted this way, 
a tree walk for a force calculation then reduces to a scanning of this
list.  If a cell must be opened, one simply examines the next node in this
list which is guaranteed to be the first child cell at the next
level of the hierarchy.  If a cell can be used,  the walk to the sibling is
simply a jump down the list to the appropriate cell labelled by the sibling
pointer.
The other main advantage of resorting  the tree nodes in this way is that
it becomes easy to prune and graft trees onto existing structures which
is essential in the parallelization of the code.

The BH tree algorithm as described above has been implemented on each
processor in the parallel code.
The node level code has been optimized
using grouping and non-recursive tree walks and improve the tree code's
efficiency considerably over simpler versions.
 
\section{Parallel Tree Codes}

The parallelization of the BH algorithm is not obvious since the
inhomogeneous distribution of particles in the oct-tree structure does not
immediately lend itself to a simple domain decomposition with load balance.
The main difficulty is that
both the particles and the tree structure
must somehow be distributed in a balanced way 
among many independent processors.
Salmon (1990) describes one possible parallelization of the tree
code which retains most of the features of the original tree code
on the level of
individual processors but introduces a new algorithm for distributing the
particles and tree structure among the processors.
We describe the algorithm below and its implementation on the CRAY T3D.
The main difference between this new code and Salmon's original code is 
the optimized implementation of the BH algorithm described in \S 2.
The parallel features are essentially the same.

%
%

\subsection{Orthogonal Recursive Bisection}

In an N-body problem,
the system of particles can be enclosed by a rectangular box
for isolated systems like individual galaxies 
or an exact cube in cosmological systems 
with periodic boundaries.
Salmon (1990) developed a parallel algorithm
for decomposing this volume into rectangular sub-volumes using the
method of orthogonal recursive bisection (ORB).
In this technique,
a volume is represented as
a hierarchy of rectangular boxes somewhat like the BH tree.
The main volume is first cut along an arbitrary dimension
at an arbitrary  position.  For a balanced tree,
the position is chosen so there are equal
numbers of particles on each side of the cut.
The resulting two volumes are cut again 
along the same or a different dimension again at an arbitrary position.
The slicing of each subvolume continues for as many levels as required.
The resulting process can be represented by 
a binary tree structure since each node points 
to two children.
An ORB tree of $n$ levels therefore results in $2^n$ subvolumes.
Most massively parallel computers are organized into partitions of $2^n$
processors ($n \le 10$ in practice) so that an $n$-level ORB tree
decomposition of an N-body volume maps conveniently onto a parallel machine.

At each level in the construction of the ORB tree, we have the freedom to
choose both the dimension and position of the volume subdivision.
For example, if the initial volume is a cube,
one can construct a ORB tree analogue of the BH tree by
cycling through the 3 dimensions and cutting only 
through the geometric center of
each subvolume.
However, with load balance in mind we wish to position our slice of
the subvolume
so that there are equal amounts of {\em computational work} on each 
side of the slice.
This is the trick introduced by Salmon to achieve load balance.
We can keep track of
the computational work for an N-body simulation by simply
summing up the number of
cell-particle interactions, the main sink of computing time.
In this way, once the particles in each subvolume are assigned to a processor,
the calculation of the acceleration should be load balanced.
Initially, the amount of computational work is unknown so it is assumed to
be the same for each particle.  The first step is therefore somewhat
load imbalanced but in practice successive steps become more balanced.

The choice of dimension for cutting the subvolumes in the
ORB tree construction is generally arbitrary but it is best to select the
longest dimension for slicing.
The tree code uses a multipole expansion to quadrupole order so we desire
rectangular domains that are as ``spherical'' as possible.

An ORB domain decomposition on a parallel machine
can be carried out by assuming the processors are layed out
in a hypercube communication network.
A machine containing $2^n$ processors can be arranged using the
communication paradigm of an $n$-dimensional hypercube.  In this
arrangement,
each processor resides at the vertex of this hypercube and can only
communicate to $n-1$ neighbors along the lines on the edge of the
hypercube.
A domain decomposition can be carried out in $n$ steps by
consecutive exchanges of particles across the $n$-dimensions of the hypercube
network.

\begin{figure}[t]
\centerline{\epsfbox{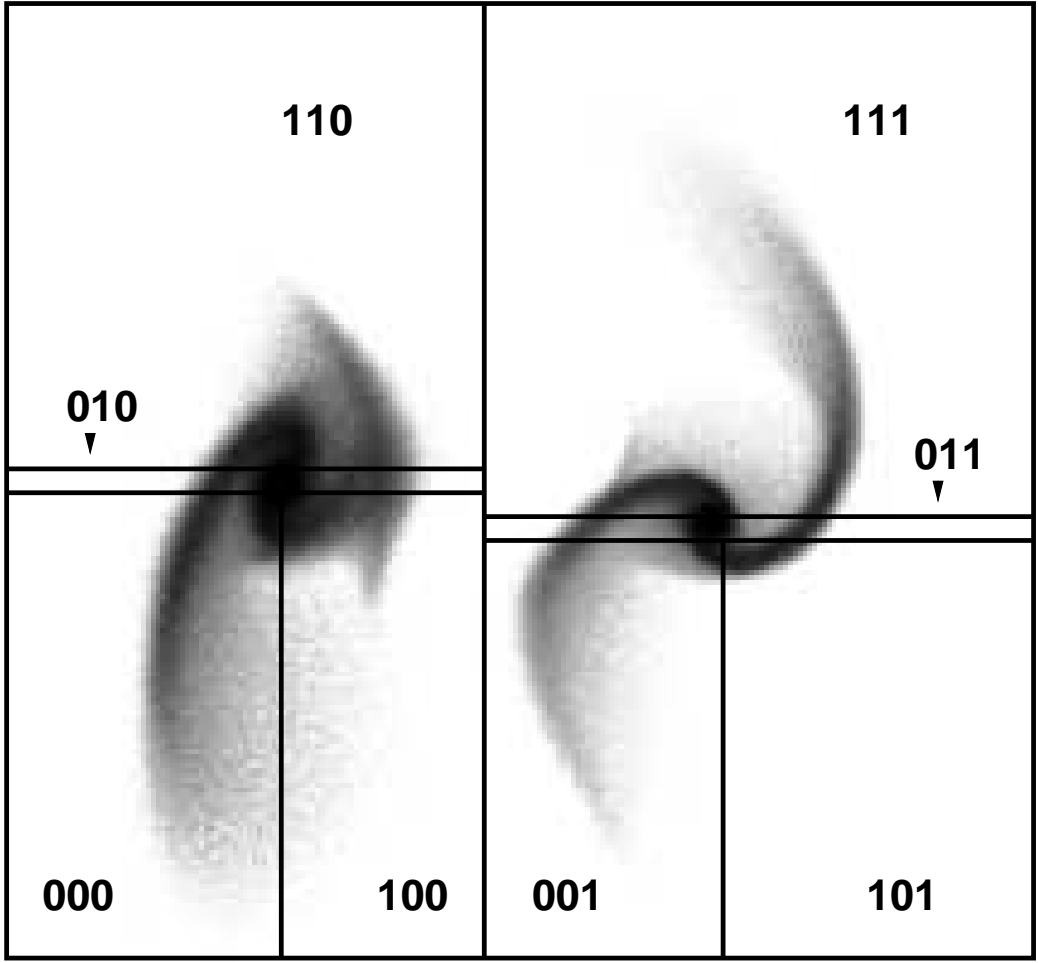}}
\caption{Domain decomposition using orthogonal recursive bisection in
a system containing 2 interacting galaxies on an 8 processor machine in 2
dimensions.
The processors are labelled with a binary number representing the 
relevant processor number in each of their
respective rectangular domains.  The domain shapes are chosen to balance
the computational load.}
\label{fig-orb}
\end{figure}

Processors are usually labeled by an integer between 0 and $2^n -1$.  
If we represent processor $i$ in binary format, its hypercube partner
across the $h$th dimension is $i\;\; {\rm XOR}\;\; 2^h$ 
where XOR is the logical
exclusive-or operation.  
This has the effect of flipping the $h$th bit from
a 0 to 1 or vice versa.
The ORB domain decomposition proceeds as follows.
A position and dimension is selected for cutting the entire domain into two
subdomains.  
For the first hypercube dimension, each processor exchanges
particles with its unique partner so that all the particles in a given
processor are on one or the other side of the first cut.
After this first iteration, half of the processors contain particles on
one side of the domain and half on the other.
At each successive iteration,
each subdomain finds the splitting position and dimension and
processors once more exchange particles as above with their corresponding
partner across the particular hypercube dimension.
Finally, each processor contains its unique
rectangular subdomain of particles (Figure \ref{fig-orb}).  
If the criterion for splitting domains
is the amount of computational work, each processor also has an allotment
of particles which will take roughly the same amount of computational
effort.
In practice,
the memory imbalance can be substantial with some processors
containing approximately twice as many particles as others although this
has been found to be tolerable.

%

\subsection{Linking the trees}

After the domain decomposition, a BH 
tree can be constructed locally in each processor using the particles
contained within its independent subvolume.
After constructing the local trees,  we would ideally
like to link them together to form the full tree describing the entire system.
\begin{figure}[t]
\centerline{\epsfbox{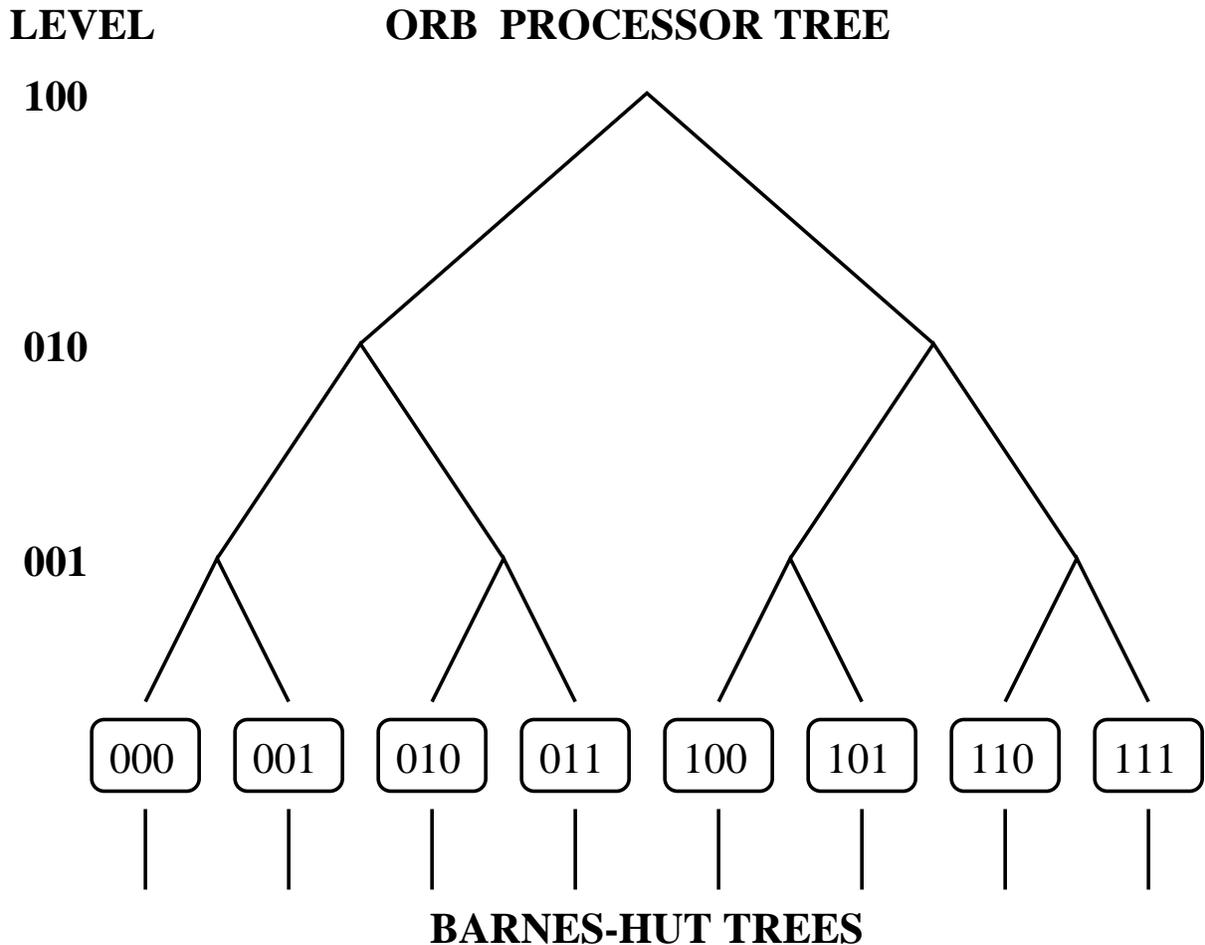}}
\caption{The hybrid tree describing the hierarchical grouping of the
particles used for the tree walk.  The top levels of the tree are the
binary structure formed by ORB domain decomposition.  The BH trees built in
each processor are grafted to the leaves of this tree to form the hybrid.
Each processor contains only pruned version of the other processor's BH
trees.}
\label{fig-kdtree}
\end{figure}
The top levels of the tree are simply the load-balanced
ORB tree structure
created by the domain decomposition, while the processors contain the
BH trees (Figure \ref{fig-kdtree}).
If a copy of this complete tree structure were available to each
processor, they could proceed independently to evaluate the forces on their
particles.  Unfortunately, the amount of memory required for a full copy of
the tree at each node is prohibitive.

Salmon (1990) solved this problem by introducing the concept of a 
{\em locally essential tree}.  Each processor does not require an entire
copy of the tree. Rather, only a significantly smaller subset of nodes
is necessary
since the BH-trees in distant volumes need only be
opened to a lesser degree for {\em all} the particles in a given
processor domain.
The opening criterion can be applied to the entire group of particles 
in a processor by calculating the distance to the nearest edge of a
processor volume in the same way
as described above in the grouping scheme used to improve the efficiency of
the  force
evaluations.  
Therefore, a processor
only needs to import significantly
pruned trees from distant processors which it can graft on to its existing
structure to create the locally essential tree.  This pruned tree of the
entire system contains all the nodes required to calculate the forces
to the required tolerance specified by the opening angle criterion.

In practice, the locally essential trees are constructed in the following
manner.  After building the local BH-tree, each processor imports the root
nodes of the trees from all of the others in the pool (Figure \ref{fig-kdtree}).
A local binary tree is built from the base up in each processor using the
imported root nodes.
A walk through the ORB tree
using a group opening-criterion determines which subset of processors 
must be examined further to gather more tree nodes if necessary.
Tree walks are performed in the needed processors using the group opening
criterion of the requesting processors.
BH tree nodes which can be used are flagged.  
The nodes of the pruned trees are
then gathered together and exported to the calling processor.
Once received the internal child and sibling links of the imported trees
are reset and the pruned
trees are grafted at the appropriate node of the ORB binary tree.
The result of this procedure is the locally essential tree which contains:
the local BH-tree, the ORB tree structure and
the trees pruned to the appropriate levels imported from other processors.

\subsection{Summary}

In summary, the parallel tree code works as follows.  
At the start, particles are
distributed randomly among the $2^n$ processors.
At each timestep, the following procedures are carried out:

\begin{enumerate}
\item ORB domain decomposition across processors.
\item Construct the local BH tree.
\item Exchange tree nodes to construct the locally essential trees.
\item Walk through trees to calculate forces.
\item Move particles along their trajectories using these forces.
\end{enumerate}

Only steps 1. and 3. require message passing.  
In both cases, messages are exchanged in a synchronous manner using the
hypercube paradigm described above.
The messages
contain arrays of
particle data (masses, positions, and velocities) 
and tree node data (masses, positions,
quadrupole moments and cell dimensions).
To ease the problems of bookkeeping, the code is written in C and particle
and tree node data are packaged in arrays of appropriate C structures.
We use the Message Passing Interface (MPI) software to handle the message
passing.  The MPI functions are supported on almost all parallel machines so
the code is portable to different platforms.
The code runs on the Cray T3D, Intel Paragon, and the IBM SP/2.

\section{Performance}

\begin{figure}[t]
\centerline{\epsfbox{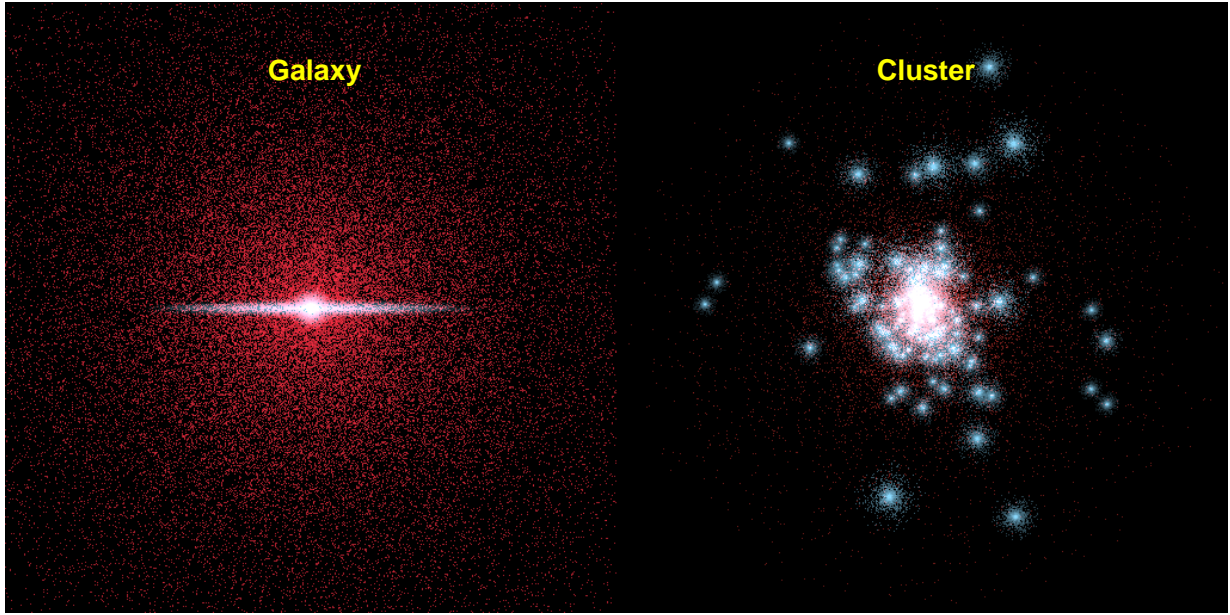}}
\caption{The galaxy and cluster models used for analyzing the performance
of the tree code.  The galaxy contains a disk, bulge and dark halo.
The cluster of galaxies contains 128 spherical galaxies and a background
cluster dark halo sampled from an Hernquist density profile.}
\label{fig-nbody}
\end{figure}

In this section,
we describe some results on the code performance on the Cray T3D in terms
of accuracy, speed and load balance.
We use two test problems to profile the code:
an isolated galaxy
composed of a disk, bulge and a halo and a spherical cluster of galaxies
following a Hernquist density profile.
The first case represents a smooth 
centrally concentrated mass distribution expected in galaxy simulations
while the second  case  exemplifies a strongly, inhomogeneous clustered 
distribution which arises in a cosmological context (Figure
\ref{fig-nbody}).

Tree codes usually run most slowly for these sorts of problems so the 
performance study here should test the code most rigorously.
More homogeneous systems run approximately twice as quickly in practice.
We note that the code has already been applied to simulations of
interacting and colliding galaxies containing $\sim$300 thousand particles 
and runs without problems for greater than 1000 timesteps
(Dubinski, Mihos, and Hernquist 1996).
The code also runs on the Intel Paragon and the IBM SP2 and note that the
performance on these machines is comparable to the T3D.

\subsection{Acceleration Errors}

\begin{figure}[t]
\centerline{\epsfbox{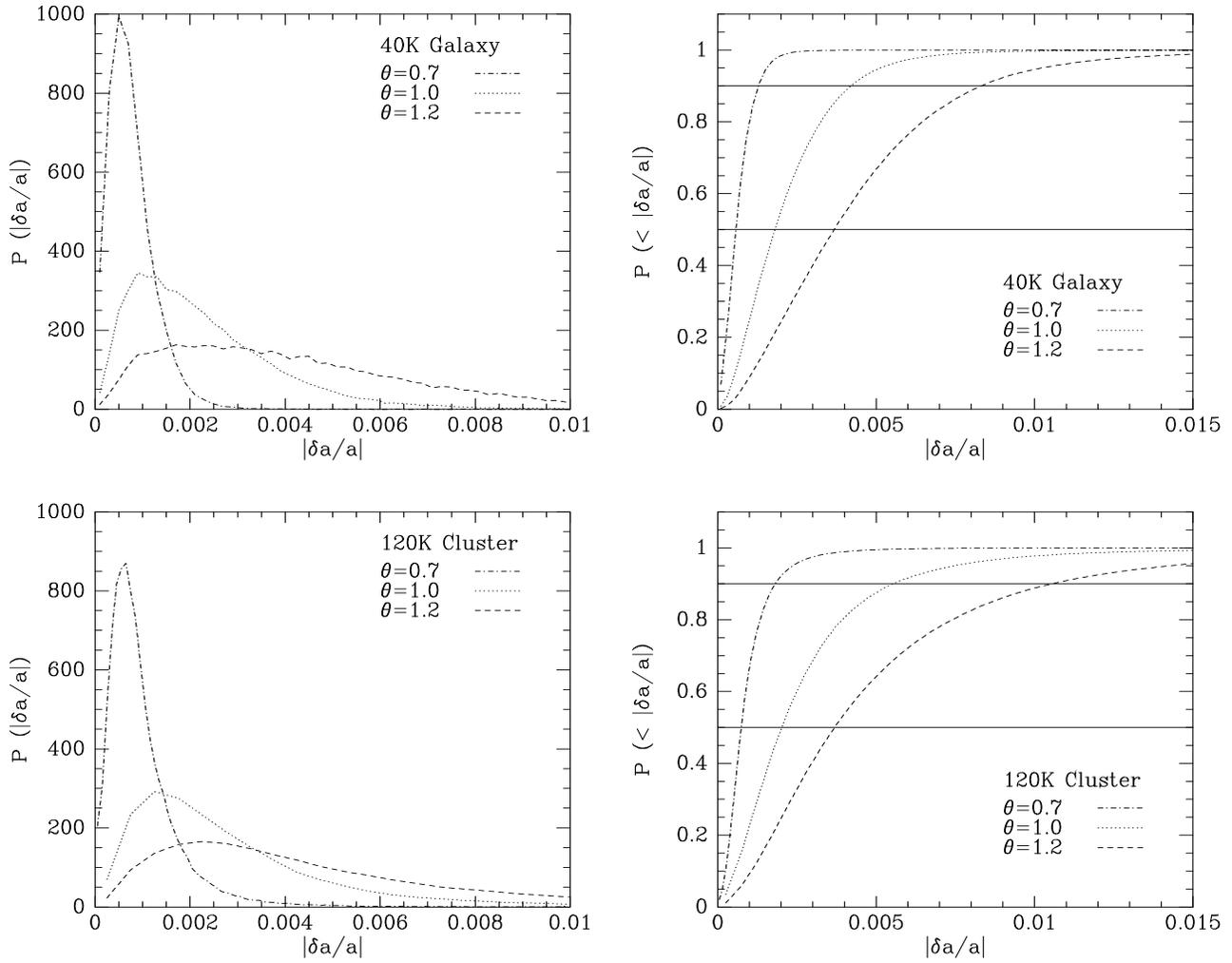}}
\caption{Distribution of acceleration errors for different particle
models using Barnes' new cell opening criterion (equation 2).  
The median error is still typically less than 1\% even for $\theta = 1.2$.}
\label{fig-err}
\end{figure}

We first measured the errors in acceleration, $|\delta {\bf a}|/|{\bf a}|$, 
for various values of the cell
opening angle, $\theta$ in single galaxy and cluster models.
We only use a single processor to measure these errors. All cell-particle
interactions are calculated to quadrupole order.
The galaxy contained 40000 particle while the cluster contained 128
galaxies of $\sim 700$ particles each plus a smooth cluster dark halo for a
total of 120000 particles.
Accelerations were first calculated using $\theta =
0.5$. Since the
errors are less than 0.02\% for this case
the accelerations are effectively exact for comparison to those
calculated with larger values of $\theta$.
Figure \ref{fig-err} shows plots of the distribution of acceleration 
errors for various
values of $\theta$ along with
the resulting cumulative distribution.

Even with $\theta=1.2$, the median error in the acceleration 
is still less than 0.5\% while 90\% of the measured accelerations are less
than 1\% for both the galaxy and the cluster.  
The addition of $\delta$ in Barnes' new opening
criterion causes many more cells to be opened than with the original
criterion and therefore allows more accurate forces for larger values
of $\theta$.  The old opening criterion gave similar error distributions 
for $\theta=1$ (Hernquist 1987).

We also found that
as the number of processors increases, 
the mean acceleration errors are found to decline while
the computing time per processor increases since
more cell openings are required.  
The top cells of the tree in a multi-processor run
created during the domain decomposition
can have large aspect ratios (Figure \ref{fig-orb}).
The result is more accurate forces than for a cubic decomposition
since the walk through the
hybrid ORB -- BH tree structure must go to deeper levels than the pure BH
tree to satisfy the cell-opening criterion.
In principle, the extra accuracy is not needed but at least the slightly
degraded performance results in a higher quality simulation.

\subsection{Memory Requirements}

The memory requirements of the parallel tree code are quite large.
The majority of the memory is assigned to the particle and cell arrays in
for the local subset of particles with additional arrays for particles and
cells imported for the locally essential trees.
The requirements for the particles and cells are respectively 15 and 20
machine words (8 bytes) with an additional arra 
for the tree nodes imported from other processors.
The memory for imported tree nodes can increase substantially for small
$\theta$ but in practice is $<$ 1 Megaword for $\theta > 1$ in simulations
that fully load the machine.
In practice, there are about 6 Mwords available on the T3D
for local particles and cells so this amounts to a maximum of 170K particles 
per processor.
The scheme used for load-balancing computing time can lead to a memory
imbalance of up to a factor of 2 in the number of particles per processor.
The practical value for simulations is therefore about $\sim 100$K
particles per processor to allow for this imbalance.
The memory is allocated and reallocated dynamically for added flexibility
in managing the changing sizes of the particle and cell lists.

\subsection{Timing}

A successful parallel algorithm should have good load
balance with a minimal overhead for processor communication.
We ran our two test cases for a few time steps using different values of
$\theta$ and different numbers of processors to examine these properties.
We used larger number of particles to fully load the machines:
640 thousand particles for the galaxy, and 1.1 million for the cluster.
We timed the computational functions (tree building/tree walking) and
communication functions (domain decomposition and tree exchange and
distribution) separately 
to measure the load balance and overall communication costs of the algorithm.

\subsubsection{Load Balance}
We define the load balance as:
\begin{equation}
B = \frac{\sum_i t_i + t_c}{{\rm max}(t_1,t_2,...,t_n) + t_c}
\end{equation}
where $t_i$ is the pure computational CPU time
for a given processor and $t_c$ is the communication time.
For this
code $t_c$ is the same for each processor since message passing is done
synchronously.
The denominator is just the elapsed wall-clock time for
each timestep.
We define the communication overhead, $C$,  as the fraction of the total
wall-clock time per step used for communication:
\begin{equation}
C = \frac{t_c}{{\rm max}(t_1,t_2,...,t_n) + t_c}
\end{equation}
The load balance and communication overhead 
for the test cases as a function of number of processors for are listed in Table 1.
We show results for the first step and third step of an integration.
In the first step,
particles are assumed to have
equal computational costs and are therefore arranged in an ORB tree with
equal numbers of particles per processor.
Load balance is in the range of 60\%-70\% while tolerable
can still be improved considerably.
We expect
this first step to be poorly load balanced since we know that in
practice particles will have a wide range of computational costs.

By the third step, the system is distributed by ORB according to computational
load.
For most cases, the load balance has improved considerably to greater than
90\% for $n<64$.
The communication overhead is generally less than 10\% of the total time
per step so the computation is being dominated by physical calculations.
For processor numbers greater than 128, there does not appear to be a major
improvement in the load balancing by the third step.  Communication costs 
become more significant and so the use of the number of cell interactions
to estimate the load is not necessarily the best way to weight the
particles.  An additional term incorporating the communication cost
in the load estimate may improve the balance.  In any case, load balance is
still better than 70\% so there is not much room left for improvement.
\begin{table}[t]
\begin{center}
\begin{tabular}{rllll}
\multicolumn{5}{c}{\sc Table 1} \\
\multicolumn{5}{c}{\sc Load Balance} \\
& \multicolumn{2}{c}{Galaxy} & \multicolumn{2}{c}{Cluster} \\ \hline
\multicolumn{1}{c}{$n$} & 
\multicolumn{1}{c}{$B$} &
\multicolumn{1}{c}{$C$} &
\multicolumn{1}{c}{$B$} &
\multicolumn{1}{c}{$C$} \\
\hline
\multicolumn{2}{l}{1st Step} &&& \\ \\
 16 & 0.77 & 0.03 & -- & -- \\
 32 & 0.71 & 0.04 & 0.85 & 0.06 \\
 64 & 0.66 & 0.09 & 0.84 & 0.12 \\
128 & 0.64 & 0.16 & 0.87 & 0.28 \\
256 & 0.69 & 0.32 & 0.87 & 0.38 \\ \hline \\
\multicolumn{2}{l}{3rd Step} &&& \\ \\
 16 & 0.95 & 0.04 & -- & -- \\
 32 & 0.95 & 0.06 & 0.93 & 0.04 \\
 64 & 0.92 & 0.09 & 0.94 & 0.06 \\
128 & 0.80 & 0.12 & 0.93 & 0.11 \\
256 & 0.73 & 0.21 & 0.88 & 0.15 \\
\hline
\end{tabular}
\end{center}
\end{table}

\subsubsection{Speed and Scaling}

\begin{figure}[t]
\epsfxsize 4.0in
\centerline{\epsfbox{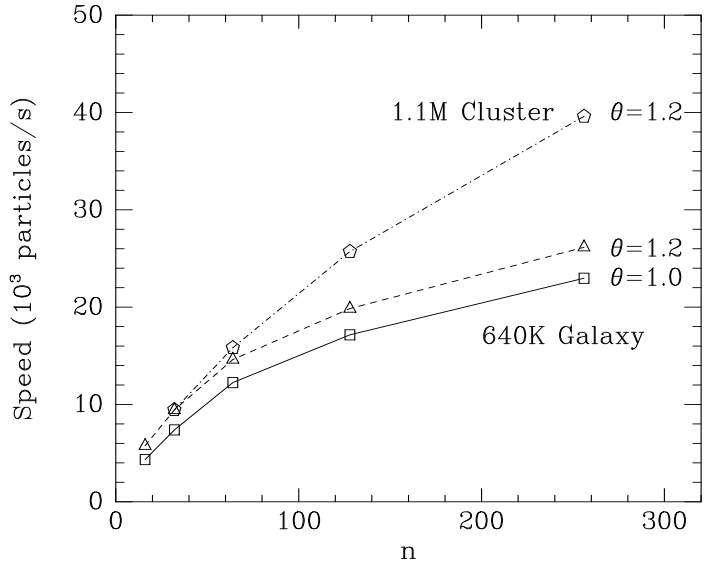}}
\caption{Code speed in particles per second.  The speed scaling 
degrades with large $n$
because of irregular domain shapes and increasing communication overhead.}
\label{fig-speed}
\end{figure}

We first measured the speed of the code in terms of the number of particles
which could be evaluated per second as a function of the number of
processors.
Speeds were calculated from the third step once the
particles had settled into a load-balanced decomposition.
The speed is simply, $S=N/t_s$, where $N$ is the number of particles and
$t_s$ is the elapsed wall-clock time per step.
Figure \ref{fig-speed} shows the code speed, $S$, for various cases.
The code shows an approximate linear scaling versus $n$ 
for $n<64$ but beyond that it falls off.  As $n$ increases,
the shapes of domains in the ORB decomposition have a larger range of
aspect ratios forcing tree walks to proceed to deeper levels to satisfy the
opening criterion.  
We found that the number of cell interactions grows with the
number of processors because of this effect.

\begin{figure}[t]
\epsfxsize 4.0in
\centerline{\epsfbox{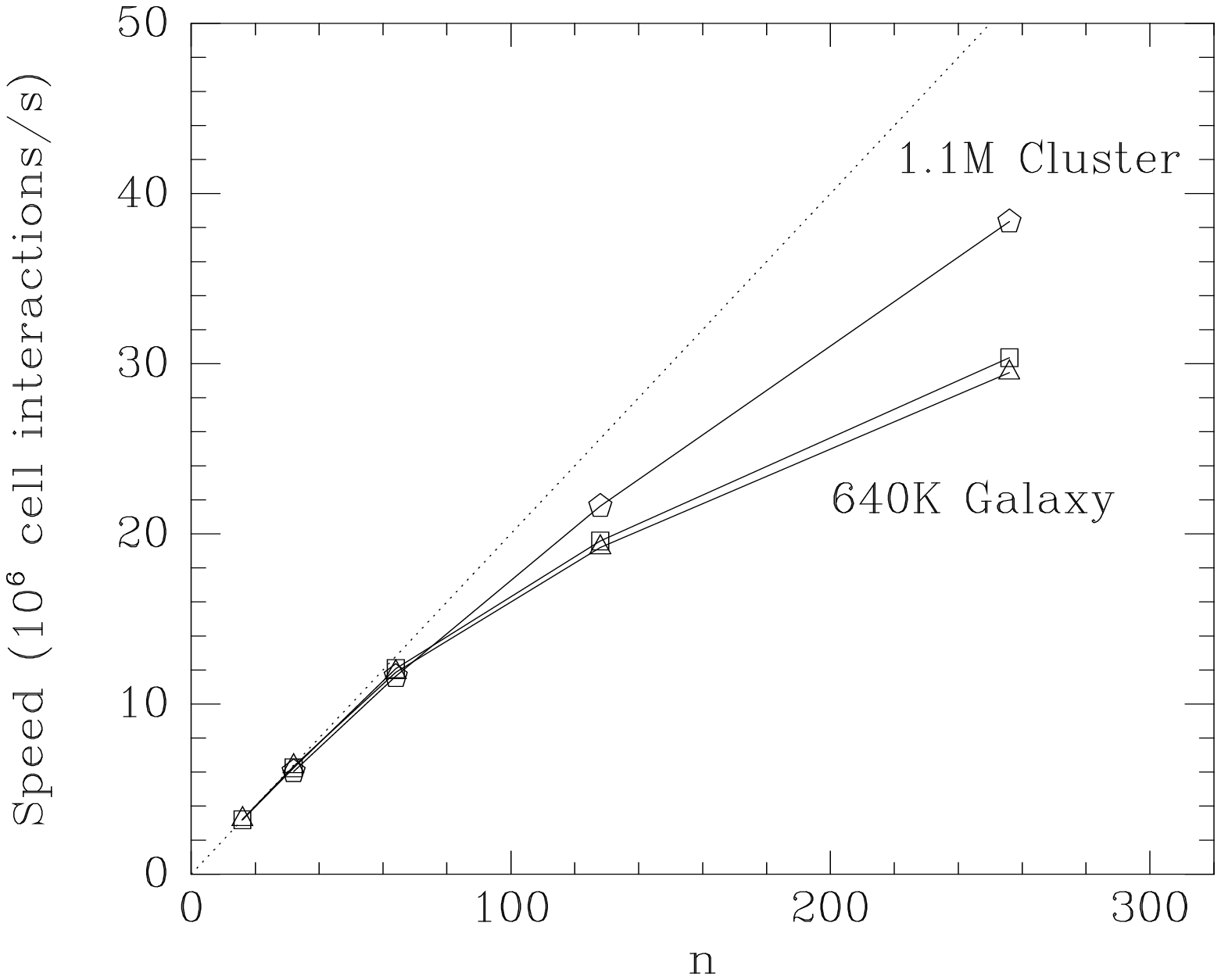}}
\caption{Code speed in cell-interactions per second, a measure of 
raw computational performance.  The dotted line shows the ideal performance
for a perfectly load-balanced algorithm. 
For $n<64$, the speed
is independent of the model showing the good load balance for these models.
For larger $n$, increasing communication costs degrade the
performance.}
\label{fig-qspeed}
\end{figure}

We also measured the speed simply as the number of cell
interactions/second, the dominant source of computation, to see how well
the raw computational speed of the code 
scales with $n$. Figure \ref{fig-qspeed} again shows a nearly linear scaling
for $n<64$ after which there is a drop off which can be attributed mainly
to the added load imbalance and communication cost for large $n$.
This speed is independent of the model for $n<64$ showing that it is an
approximate measure of the raw computational speed.
There are 72 floating point calculations per quadrupole order interaction
plus one square root.  There is also some computation associated with the
tree walk and some direct particle-particle interactions.  There are
therefore approximately 100 floating point operations of useful work per
cell interaction.  The code speed on the T3D is therefore in the range of 15-20
MFlops/processor depending on $n$.  
Similar estimates are found using the
``Apprentice'' profiling utility on the T3D.
This value falls short of the peak theoretical speed of 150 MFlops/node.
The deficit results from the large of amount of bookkeeping needed in
shuffling and resorting particles and tree nodes at various times before
actually doing the useful work of computing the forces. 

In practice, the code running on a 16 node partition performs almost as well 
as Hernquist's (1990) vectorized tree code on the CRAY C90.
Walker's 500,000 particle of satellite accretion ran at a rate of $\sim$4000
particles per second with $\theta=0.7$ (Walker, Mihos \& Hernquist 1996).
Parallel treecodes running on the newest parallel machines 
therefore offer about a factor of 10 fold increase
in problem size or speed over vectorized tree codes.

We also note that the speed of this parallel code is comparable to Warren \&
Salmon's (1994) when comparing their homogeneous sphere benchmarks.

\section{Conclusions}

A parallel tree code based on Salmon's algorithm has been 
implemented  using the MPI message passing software
on 3 parallel machines: the Cray T3D, Paragon
and the  IBM SP/2.
In principle, the code can also run on a 
small network of workstations although we have not had the opportunity to
test it this way.
The code incorporates an improved version of the BH tree algorithm
including particle grouping for tree walks, 
non-recursive tree walking and a ``safe'' cell opening
criterion.
The code has been applied successfully to simulations of galaxy interactions
(Dubinski, Mihos, \& Hernquist 1996)
and is currently being applied to a variety of other problems in galaxy
dynamics.

The code works best with a small number of processors that 
are heavily loaded with particles.
If $N < 10^6$, we find that  the number of processors should be less than 64
to get the best use of computing time.
Bigger problems will of course require the added memory 
of more processors.

This code's main disadvantage is its excessive demand for memory.
The memory needs
become excessive when $\theta$ becomes too small.
The use of locally essential trees is the main culprit in this respect.
Perhaps, this method of load balancing can be replaced by an
asynchronous message passing scheme in which tree walks in various
processors are done on demand by distributing particles among the
processors.
At present, the code
is still marginally time limited and available memory will be greater
in the next generation of machines.

We are now in the process of adapting the code for both cosmological
simulations with periodic boundary conditions and smoothed particle
hydrodynamics (Dave, Dubinski \& Hernquist 1996).
With this new code, we should be able to increase the typical number of
particles used in current CRAY C90 TREESPH (Hernquist \& Katz 1989)
simulation by a factor of 10-100.  
The improved dynamic range should allow simulations
of galaxy formation in a full cosmological context.

\acknowledgments
I would like to thank Romeel Dave, Uffe Hellsten, Lars Hernquist and
Guohong Xu for useful comments.
I acknowledge grants for supercomputing time at the
National Center for Supercomputing Applications, the Pittsburgh
Supercomputing Center, the San Diego Supercomputing Center and the Cornell
Theory Center.

\clearpage
\references
\refpar Appel, A.W. 1985, In SIAM Journal  on Scientific and Statistical Computing, 6, 85 
\refpar Barnes, J. \& Hut P. 1986, Nature, 324, 446
\refpar Barnes, J.E 1990, J. of Comp. Phys., 87, 161 
\refpar Barnes, J.E. 1994, In Computational Astrophysics, Eds. J. Barnes et al. (Springer-Verlag: Berlin)
\refpar Benz, W., Bowers, R.L., Cammeron, A.G.W. \& Press, W.H. 1990, ApJ, 348, 647
\refpar Bhatt, S.; Chen, M.; Lin, C.-Y.; Liu, P. 1992, In Proceedings Scalable High Performance Computing Conference (Los Alimitos: IEEE Comput. Soc. Press)
\refpar Dave, R., Dubinski, J. \& Hernquist, L. 1996, in preparation
\refpar Dikaiakos, M.D. \& Stadel, J. 1995, preprint
\refpar Dubinski, J. 1988, M.Sc. Thesis, University of Toronto
\refpar Dubinski, J., Mihos, J.C, \& Hernquist, L. 1996, ApJ, in press
\refpar Fox, G.C., Williams, R.D, \& Messina, P.C. 1994, Parallel Computing Works!, San Francisco, CA: Morgan Kaufmann Publishers Inc.
\refpar Hernquist, L \& Katz, N. 1989, ApJS, 70, 419
\refpar Hernquist, L. 1987, ApJSS,, 64, 715
\refpar Hernquist, L. 1990, J. Comp. Phys., 87, 137
\refpar Hernquist, L., Bouchet, F.R., \& Suto, Y. 1991, ApJSS, 75, 231 
\refpar Hillis, W.D. \& Barnes, J. 1987, Nature, 326, 27
\refpar Jernigan, J. G. \& Porter, D. H 1989, ApJS, 71, 871
\refpar Makino, J. \& Hut, P. 1989, Comput. Phys. Rep., 9, 199 
\refpar Pal Singh, J. 1993, Ph. D. Thesis, Stanford University
\refpar Pal Singh, J., Holt, C., Totsuka, T., Gupta, A., \& Hennessy, J. 1995, Journal of Parallel and Distributed Computing, 27, 118.
\refpar Salmon, J. 1990, Ph.D. Thesis, "Parallel Hierarchical N-body Methods",     California Institute of Technology. 
\refpar Salmon, J.K. \& Warren, M.S. 1994, Journal of Computational Physics, 111, 136 
\refpar Salmon, J.K., \& Warren, M.S. 1994, International Journal of Supercomputing  Applications, 8, 129 
\refpar Steinmetz, M. \& Muller, E. 1993, AA, 268, 391
\refpar Walker, I., Mihos, J.C. \& Hernquist, L. 1996, ApJ, in press
\refpar Warren, M., and Salmon J. 1992, ApJ, 401, 
\refpar Warren, M.S. 1994, Ph.D Thesis, Univeristy of California, Santa Barbara 
\refpar Warren, M.S. \& Salmon, J.K. 1993, In Proc. of Supercomputing '93, 12 
\endreferences
\end{document}